\documentstyle[b_phys,12pt]{article}
\def\beq{\begin{equation}}
\def\eeq{\end{equation}}
\def\sss{\scriptscriptstyle}
\def\cp{{\sss CP}}
\def\vub{V_{ub}}
\def\vtd{V_{td}}
\def\bbar{{\overline B}}
\def\fbar{{\bar f}}
\def\ks{K_{\sss S}}
\def\dcp{D_\cp}
\def\dbar{{\overline{D^0}}}
\def\roughly#1{\mathrel{\raise.3ex\hbox{$#1$\kern-.75em\lower1ex\hbox{$\sim$}}}}
\def\lsim{\roughly<}

\def\Wolf{{1}}
\def\Bigietal{{2}}
\def\PASCOS{{3}}
\def\bwidth{{4}}
\def\selftag{{5}}
\def\nonCPeigenstate{{6}}
\def\isospin{{7}}
\def\Bsmixing{{8}}
\def\refnine{{9}}
\def\bsdkref{{10}}
\def\bsdphiref{{11}}
\def\GroWyler{{12}}
\def\Besson{{13}}
\def\dunietz{{14}}
\def\isicomment{{15}}
\def\Psiphi{{16}}
\def\Gronauhelp{{17}}
\def\PsiKstar{{18}}
\def\AKL{{19}}

\rightline{DAPNIA/SPP 93-23}
\rightline{NSF-PT-93-4}
\rightline{UdeM-LPN-TH-93-184}

\centerline{{\large\bf In Pursuit of Gamma}\footnote{Talk given by B.
Kayser at the {\it Workshop on B Physics at Hadron Accelerators}, Snowmass,
Colorado, June 21-July 2, 1993.}}

\centerline{R. Aleksan}
\centerline{\em DAPNIA, CE Saclay}
\centerline{\em F-91191 Gif-sur-Yvette Cedex, France}
\centerline{B. Kayser}
\centerline{\em Division of Physics, National Science Foundation}
\centerline{\em 4201 Wilson Blvd., Arlington, VA 22230 USA}
\centerline{and}
\centerline{D. London}
\centerline{\em Laboratoire de physique nucl\'eaire, Universit\'e de
Montr\'eal}
\centerline{\em C.P. 6128, succ. A, Montr\'eal, Qu\'ebec, CANADA, H3C 3J7}

\begin{document}
 \pagestyle{plain}    

\vskip3truemm
\centerline{\bf ABSTRACT}

After reviewing techniques for extracting clean information on CP-violating
phase angles from $B$ decays, we explain the rules for finding decay modes
that can probe the phase angle $\gamma$ of the unitarity triangle. We
identify the more promising of these ``$\gamma$ modes,'' estimate their
branching ratios, and examine the degree to which they are theoretically
clean. We then show that when the quark mixing matrix is not approximated
as usual, but is treated exactly, none of the ``$\gamma$ modes'' actually
measures $\gamma$. Rather, each of them measures $\gamma$ plus some
correction. In all modes, the correction is small enough to be disregarded
in first-generation experiments, but in some of them, it may be large
enough to be observed in second-generation experiments.

Our treatment of the $\gamma$ modes calls attention to the fact that when
the quark mixing matrix is treated exactly, there are six unitarity
triangles, rather than just one triangle. However, only four of the angles
in these six triangles are independent. Examining the role played by these
four angles, we discover that, in principle at least, measurements of
nothing but CP-violating asymmetries in $B$ decays are sufficient to
determine the entire quark mixing matrix.

\onehead{1.}{INTRODUCTION}         

In the Standard Model, CP violation is caused by complex elements in the
unitary C(abibbo)-K(obayashi)-M(askawa) quark mixing matrix, $V$. If this
Model is correct, then in many $B$ decays there should be large
CP-violating asymmetries from which theoretically clean information on the
phases in $V$ can be extracted. This information can then be used to
confirm in detail that phases in $V$ really are the origin of CP violation,
or to exhibit an inconsistency of the theory.

\newpage
In the Wolfenstein approximation$^\Wolf$ to $V$, there is a phase
convention in which this matrix is real except for the elements $\vub$ and
$\vtd$. In this approximation, experiments on CP violation in $B$ decays
are usually described as probes of the angles $\alpha$, $\beta$ and
$\gamma$ of the ``unitarity triangle,'' shown in Fig.~1. That the legs in
this Figure form a closed triangle follows from the unitarity constraint
that the $d$ and $b$ columns of $V$ must be orthogonal, and the assumption
that there are only three generations. From Fig.~1 we see that in the
Wolfenstein approximation,
\beq
\gamma\simeq - arg(\vub) ~,~~~~~ \beta\simeq - arg(\vtd) ~,~~~~~
\alpha\simeq \pi + arg(\vub) + arg(\vtd).
\eeq
Thus, probing the angle $\gamma$ -- the focus of the Gamma Working Group
within this Snowmass Workshop -- amounts to probing the phase of $\vub$.

\begin{figure}
\vspace{50mm}
\caption{The unitarity triangle expressing orthogonality of the $d$ and $b$
columns of the CKM matrix.}
\end{figure}

In Sec.~2, we recall what must be measured to extract clean CKM phase
information from the decays of {\it neutral} $B$ mesons. We then identify a
number of $B$ decay modes which potentially can yield the phase of $\vub$.
We estimate the branching ratios, and comment on the degree of theoretical
cleanliness, of these modes.

In Sec.~3, we ask what happens if one treats the CKM matrix exactly, rather
than in the Wolfenstein approximation. The single unitarity triangle of
Fig.~1 is then replaced by the six unitarity triangles of Fig.~9, which
express the orthogonality of any pair of columns, or any pair of rows, of
$V$. We find that when the Wolfenstein approximation is not made, {\it
none} of the $B$ decays modes proposed so far as ``probes of the angle
$\gamma$'' actually measure $\gamma$. Instead, each of these modes measures
$\gamma$ plus corrections which are small angles in the triangles other
than the one of Fig.~1. We explore the degree to which these corrections
may undermine the interpretation of these decay modes as probes of
$\gamma$.

In Sec.~4, we report on a general analysis of the unitarity triangles and
CP-violating phases when the CKM matrix is treated exactly. We find that
the CKM phase yielded by a theoretically-clean decay mode is always a
simple linear combination of angles in the unitarity triangles. Moreover,
at least in principle, measurements of CP-violating asymmetries in $B$
decays are sufficient to determine, not only some angles in one unitarity
triangle, but the entire CKM matrix.

\newpage
\onehead{2.}{POTENTIAL PROBES OF GAMMA}

\ttwohead{2.1}{Extraction of CKM Phases}

In general, {\it theoretically clean} CKM phase information can be
extracted only from the decays of the {\it neutral} $B$ mesons, $B_d$ and
$B_s$. In the decays of either of these mesons, we are usually interested
in some final state which can come both from the pure $B$ and from the pure
$\bbar$. Now, owing to $B$-$\bbar$ mixing, a particle born at time $t=0$ as
a pure $\vert B_q \rangle$, $q=d$ or $s$, evolves in time $t$ into a state
$\vert B_q(t) \rangle$ which is a linear superposition of $\vert B_q
\rangle$ and $\vert \bbar_q \rangle$.$^{\Bigietal,\PASCOS}$ In the
(excellent) approximation that $B$-$\bbar$ mixing is dominated by a
$t$-quark box diagram, this superposition is given by
\beq
\vert B_q(t) \rangle = exp\left(-i(m_q-i{\Gamma_q\over 2})t\right) \left[
c_q \vert B_q \rangle + i \, \omega_q s_q \vert \bbar_q \rangle \right].
\label{btimedep}
\eeq
Here, $m_q$ is the average mass of the two mass eigenstates of the
$B_q$-$\bbar_q$ system, and $\Gamma_q$ is their common width.$^\bwidth$
With $\Delta m_q$ their mass difference,
\beq
c_q \equiv \cos\left({\Delta m_q\over 2} \, t\right),~{\rm and}~
s_q \equiv \sin\left({\Delta m_q\over 2} \, t\right).
\eeq
Finally, $\omega_q$ is the CKM phase of the amplitude $A(B_q\to\bbar_q)$
for $\vert B_q \rangle \to \vert \bbar_q \rangle$, and is given by
\beq
\omega_q = {V_{tq}V_{tb}^* \over V_{tq}^*V_{tb}}~.
\label{omegaq}
\eeq
In a similar fashion, a particle born at $t=0$ as a pure $\vert \bbar_q
\rangle$ evolves in time $t$ into a linear superposition $\vert \bbar_q(t)
\rangle$ of $\vert \bbar_q \rangle$ and $\vert B_q \rangle$ given by an
expression analogous to that of Eq.~(\ref{btimedep}). Suppose, now, that
$f$ is some final state which can come both from a pure $B_q$ and from a
pure $\bbar_q$. From Eq.~(\ref{btimedep}), the amplitude $A(B_q(t)\to f)$
for the meson $B_q(t)$ which at time $t=0$ was a pure $B_q$ to decay into
$f$ at time $t$ is
\beq
A(B_q(t)\to f) = exp\left(-i(m_q-i{\Gamma_q\over 2})t\right) \left[
c_q \, A( B_q \to f) + i \, \omega_q s_q \, A(\bbar_q \to f)  \right].
\label{Binterference}
\eeq
The corresponding time-dependent decay rate, $\Gamma_{q,f}(t)\equiv \vert
A(B_q(t)\to f)\vert^2$, then contains a term representing the interference
between the $A(B_q\to f)$ and $A(\bbar_q\to f)$ terms in
Eq.~(\ref{Binterference}).

Let us now turn to the CP-mirror-image process $\bbar_q(t)\to\fbar$, in
which the meson $\bbar_q(t)$ born at $t=0$ as a pure $\bbar_q$ decays into
the final state $\fbar$, the CP-mirror-image of $f$. The rate for this
process, ${\overline\Gamma}_{q,\fbar}(t) \equiv \vert A(\bbar_q(t)\to
\fbar)\vert^2$, also contains an $A(B_q\to f)$-$A(\bbar_q\to f)$
interference term. However, when the CKM matrix elements are complex, this
interference term has, in general, a different magnitude than its
counterpart in $\Gamma_{q,f}(t)$. This difference leads to a CP-violating
difference between ${\overline\Gamma}_{q,\fbar}(t)$ and $\Gamma_{q,f}(t)$,
which one would like to observe. In order to observe it, one must know in
each event whether the decaying meson was born as a $B_q$ or a $\bbar_q$.
That is, one must tag it as one of these by observing a flavor-revealing
decay of an accompanying beautiful meson or baryon. It may also be possible
to use an interesting, recently-proposed ``self-tagging''
method.$^\selftag$

Suppose that the final state $f$ is a CP eigenstate, so that $\fbar$ is the
same as $f$. If $f$ has intrinsic CP parity $\eta_f$, the decay rates
$\Gamma_{q,f}(t)$ and ${\overline\Gamma}_{q,\fbar}(t) =
{\overline\Gamma}_{q,f}(t)$ are given by$^\Bigietal$
\begin{eqnarray}
\Gamma_{q,f}(t) & \propto & exp\left(-\Gamma_q t\right)
\left[ 1 + \eta_f \sin \varphi_{q,f} \sin(\Delta m_q t) \right],
\nonumber\\
{\overline\Gamma}_{q,f}(t) & \propto & exp\left(-\Gamma_q t\right)
\left[ 1 - \eta_f \sin \varphi_{q,f} \sin(\Delta m_q t) \right].
\label{CPrates}
\end{eqnarray}
Here, $\varphi_{q,f}$ is the phase of some product of CKM elements whose
identities depend on $q$ and $f$. It is $\varphi_{q,f}$ that we would like
to determine from the asymmetry in the decay rates (\ref{CPrates}). We
shall be interested in decays where $\varphi_{q,f}$ is $\gamma$, or perhaps
$2\gamma$.

Suppose, next, that the final state $f$ is not a CP eigenstate, but has a
CP conjugate $\fbar$ distinct from itself. An example of interest is
$f=D_s^+ K^-$, $\fbar=D_s^- K^+$. Theoretically clean CKM phase information
can still be extracted.$^\nonCPeigenstate$ There are now four decay rates
which can be measured. They are given by
\begin{eqnarray}
\Gamma_{q,f}(t) & = & exp\left(-\Gamma_q t\right)
\left[ c_q^2 \, M^2_{q,f} + s_q^2 \, {\overline M}^2_{q,f} + 2 c_q s_q \,
M_{q,f} {\overline M}_{q,f} \sin (\varphi_{q,f} + \theta_{q,f}) \right]
\nonumber\\
{\overline\Gamma}_{q,\fbar}(t) & = & exp\left(-\Gamma_q t\right) \left[
c_q^2 \, M^2_{q,f} + s_q^2 \, {\overline M}^2_{q,f} + 2 c_q s_q \, M_{q,f}
{\overline M}_{q,f} \sin ( -\varphi_{q,f} + \theta_{q,f}) \right]
\nonumber\\
{\overline\Gamma}_{q,f}(t) & = & exp\left(-\Gamma_q t\right) \left[ c_q^2
\, {\overline M}^2_{q,f} + s_q^2 \, M^2_{q,f} - 2 c_q s_q \, M_{q,f}
{\overline M}_{q,f} \sin ( \varphi_{q,f} + \theta_{q,f}) \right] \nonumber\\
\Gamma_{q,\fbar}(t) & = & exp\left(-\Gamma_q t\right) \left[ c_q^2
\, {\overline M}^2_{q,f} + s_q^2 \, M^2_{q,f} - 2 c_q s_q \, M_{q,f}
{\overline M}_{q,f} \sin ( -\varphi_{q,f} + \theta_{q,f}) \right].
\label{nonCPrates}
\end{eqnarray}
Here, $\Gamma_{q,f}(t)$ is the rate for decay of $B_q(t)$ into $f$,
${\overline\Gamma}_{q,\fbar}(t)$ is the rate for decay of $\bbar_q(t)$ into
$\fbar$, etc. The angle $\varphi_{q,f}$ is, as before, the phase of some
product of CKM elements whose identities depend on $q$ and $f$. As before,
$\varphi_{q,f}$ is the quantity we would like to determine, and we shall be
interested here in decays where $\varphi_{q,f}$ is $\gamma$ or $2\gamma$.
The constants $M_{q,f}$ and ${\overline M}_{q,f}$ are, respectively, the
magnitudes of the amplitudes $A(B_q\to f)$ and $A(\bbar_q\to f)$. It is
desirable that $M_{q,f}$ and ${\overline M}_{q,f}$ be comparable, so that
the rates (\ref{nonCPrates}) will be sensitive to $\varphi_{q,f}$. Finally,
$\theta_{q,f}$ is a strong-interaction phase.

With $\Gamma_q$ and $\Delta m_q$ known, measuring the decay rates
(\ref{nonCPrates}) more than suffices to determine the quantities
$s_\pm(q,f)\equiv \sin(\pm \varphi_{q,f} + \theta_{q,f})$. In turn, these
quantities determine $\sin^2 \varphi_{q,f}$, up to a two-fold ambiguity,
via the expression
\beq
\sin^2 \varphi_{q,f} = {1\over 2} \left[ 1 - s_+(q,f) s_-(q,f) \pm
\sqrt{ ( 1 - s_+^2(q,f)) ( 1 - s_-^2(q,f)) } \right].
\eeq
If, contrary to what we have assumed here, the two mass eigenstates of the
$B_q$-$\bbar_q$ system have widths which differ enough to result in
measurable effects, it becomes possible to experimentally resolve some of
the ambiguities in the determination of $\varphi_{q,f}$.

The decay rates (\ref{CPrates}) and (\ref{nonCPrates}) hold when $A(B_q\to
f)$ and $A(\bbar_q\to f)$ are each dominated by a single Feynman diagram,
so that they each have a well-defined CKM phase. When, instead, $A(B_q\to
f)$ receives significant contributions from several Feynman diagrams with
different CKM phases, the extraction of clean CKM phase information from
experimental decay rates is either impossible or requires measurement of
rates for several isospin-related decays.$^\isospin$ When several diagrams
contribute significantly, the largest one is usually a tree diagram, and
the others are usually penguin diagrams. In exploring the usefulness of
each decay mode proposed as a probe of the angle $\gamma$, we will consider
the degree to which penguin or other diagrams with CKM phases different
from that of the dominant diagram might contribute significantly to the
mode.

In the decay rates (\ref{CPrates}) and (\ref{nonCPrates}), the violation of
CP invariance, and the information on the CKM phase $\varphi_{q,f}$
producing this violation, are in the term proportional to $2 c_q s_q =
\sin(\Delta m_q t)$. To learn about $\varphi_{q,f}$, one would like to
measure the time dependence of the rates and uncover this term. When $f$ is
not a CP eigenstate and the rates are described by Eqs.~(\ref{nonCPrates}),
measurement of their time dependence is absolutely essential. Not all four
decay rates need be measured. Indeed, it is easy to see$^\PASCOS$
that, say, $\Gamma_{q,f}(t)$ and ${\overline\Gamma}_{q,\fbar}(t)$ alone
suffice to determine $\sin^2 \varphi_{q,f}$. However, if we measure only
the time integrals of the rates (\ref{nonCPrates}), then we cannot
determine $\sin^2 \varphi_{q,f}$, even if we measure the time integrals of
all four of the rates. For, if Eqs.~(\ref{nonCPrates}) hold, then clearly
we must have
\beq
\Gamma_{q,f}(t) + {\overline\Gamma}_{q,f}(t) =
\Gamma_{q,\fbar}(t) + {\overline\Gamma}_{q,\fbar}(t)~.
\eeq
Now, when the decay rates in this constraint are replaced by their time
integrals, they become merely four numbers, instead of four functions of
time, and the constraint implies that only {\it three} of these four
numbers are independent. But the decay rates (\ref{nonCPrates}), and their
time-integrated analogues, depend on {\it four} unknowns: $M_{q,f}$,
${\overline M}_{q,f}$, $\varphi_{q,f}$, and $\theta_{q,f}$. Hence, it is
impossible to determine $\varphi_{q,f}$ from the time-integrated rates.
When $f$ is a CP eigenstate and the decay rates are described by
Eqs.~(\ref{CPrates}), then in principle one can extract $\sin
\varphi_{q,f}$ from a knowledge of the time-integrated rates alone.
However, in the case of $B_s$ decay, this will be extremely difficult if,
as we expect,$^\Bsmixing$ $\Delta m_s$ is an order of magnitude larger than
$\Gamma_s$. When $x_s\equiv \Delta m_s/\Gamma_s$ is large, the fractional
contribution of the CP-violating $\sin \varphi_{q,f} \sin(\Delta m_s t)$
term in Eqs.~(\ref{CPrates}) to the decay rate gets reduced by a factor of
$\sim 1/x_s$ when the rate is integrated over time.

In view of these circumstances, we assume here that when one is seeking to
extract CKM phase information from a neutral $B$ decay, the time dependence
of the decay rate must be measured, except perhaps in $B_d$ decay to a CP
eigenstate.

\ttwohead{2.2}{Neutral $B$ Decay Modes That Can Probe Gamma}

In which neutral $B$ decays can we identify the CKM phase $\varphi_{q,f}$
that is probed as $\gamma$ or $2\gamma$? As we have discussed, the CP
violation that we study in the decay $B_q(t)\to f$ results from
interference between the two terms in the decay amplitude
(\ref{Binterference}). The CKM phase $\varphi_{q,f}$ that is probed by
$B_q(t)\to f$ is, therefore, just the relative CKM phase of these two
terms. Thus, remembering that $\omega_q$ is the CKM phase of $A(B_q\to
\bbar_q)$, we see from Eq.~(\ref{Binterference}) that the $\varphi_{q,f}$
probed by $B_q(t)\to f$ is given by
\beq
\varphi_{q,f} = CKM~Phase~\left[
{ A(B_q\to f) \over A(B_q\to\bbar_q) \times A(\bbar_q\to f) } \right].
\label{CKMphase}
\eeq
Instead of referring to Eq.~(\ref{Binterference}), we may think of the CP
violation in $B_q(t)\to f$ as resulting from interference between the
amplitude $A(B_q\to f)$ for the particle born as a pure $B_q$ to decay
directly to $f$, and the amplitude $A(B_q\to\bbar_q) \times A(\bbar_q\to
f)$ for this particle to convert, via mixing, into a $\bbar_q$ which then
decays into $f$. Once again we conclude that the $\varphi_{q,f}$ probed by
$B_q(t)\to f$ is given by (\ref{CKMphase}).

Now, recall that in the Wolfenstein approximation to the CKM matrix, all
CKM elements are real save $\vub$ and $\vtd$, and $\gamma=-arg(\vub)$. In
this approximation, the CKM phase of $A(B_d\to\bbar_d)$ is
\beq
arg(\vtd/V_{td}^*) = - 2 \beta~,
\eeq
while that of $A(B_s\to\bbar_s)$ is
\beq
arg(V_{ts}/V_{ts}^*) = 0~.
\eeq
Thus, from Eq.~(\ref{CKMphase}), we can probe $\gamma$ by studying $B_s(t)$
decays in which the phase of \\
$A(B_s\to f)/A(\bbar_s\to f)$ is essentially $\gamma$. This will be the
case when each of $B_s\to f$ and $\bbar_s\to f$ is dominated by a tree
diagram, and either (a) the tree diagram for $B_s\to f$ involves one of the
processes
\beq
{\bar b} \to {\bar u} + \left\{\matrix{ c{\bar s} \cr c{\bar d} \cr
u{\bar s} \cr u{\bar d} \cr}\right.~,
\label{bdecay}
\eeq
or (b) the tree diagram for $\bbar_s\to f$ involves one of the processes
\beq
b \to u + \left\{\matrix{ {\bar c} s \cr {\bar c} d \cr
{\bar u} s \cr {\bar u} d \cr}\right.~,
\label{bbardecay}
\eeq
or both. When both (\ref{bdecay}) and (\ref{bbardecay}) are involved, the
CKM phase of $A(B_s\to f)/A(\bbar_s\to f)$ is obviously
$arg(V_{ub}^*/\vub)=2\gamma$. When only one of them is involved, the other
is replaced by a (real) $b\to c$ or ${\bar b} \to {\bar c}$ transition, so
that the phase of $A(B_s\to f)/A(\bbar_s\to f)$ is $\gamma$.

We have considered the hadronic $B_s$ decay modes produced by tree diagrams
for the quark processes (\ref{bdecay},\ref{bbardecay}). We have tried to
identify the modes that have advantageous branching ratios, and in which
the interfering decay amplitudes $A(B_s\to f)$ and $A(\bbar_s\to f)$ are
each dominated by a single Feynman diagram$^\refnine$ and have comparable
magnitudes. The most promising modes we found are listed, together with
their estimated branching ratios, in Table 1. These branching ratios were
obtained by comparing the modes of interest to others whose branching
ratios are already known. We now discuss the modes in Table 1 in turn.

\begin{table}
\hfil
\vbox{\offinterlineskip
\halign{&\vrule#&
   \strut\null~\hfil#\hfil\null~\cr
\noalign{\hrule}
height2pt&\omit&&\omit&\cr
& Decay Mode && Branching Ratio & \cr
height2pt&\omit&&\omit&\cr
\noalign{\hrule}
height5pt&\omit&&\omit&\cr
& $B_s\to D_s^\pm K^\mp$ && $2 \times 10^{-4}$ & \cr
height5pt&\omit&&\omit&\cr
& $B_s\to D^0\phi,\dbar\phi$ && $2 \times 10^{-5}$ & \cr
height5pt&\omit&&\omit&\cr
& $B_s\to \rho^0 \ks$ && $5 \times 10^{-7}$ & \cr
height5pt&\omit&&\omit&\cr
\noalign{\hrule}}}
\caption{$B_s$ decay modes that can probe the angle $\gamma$.}
\end{table}

\noindent
$\bullet$ $B_s(t),\bbar_s(t)\to D_s^\pm K^\mp$:$^\bsdkref$

Here the final state $f\equiv D_s^+ K^-$ is distinct from its CP conjugate,
$\fbar \equiv D_s^- K^+$, and one uses the expressions (\ref{nonCPrates})
to analyze the time-dependent decays of $B_s(t)$ and $\bbar_s(t)$ into $f$
and $\fbar$. Tagging of the parent meson and study of the time dependence
of the decays are essential.

The expected branching ratio is relatively large. The value quoted in Table
1, $2\times 10^{-4}$, is for the decay $B_s\to D_s^- K^+$. Like all the
values quoted, it assumes the parent to be a pure $B_s$ and neglects
mixing. The value $2\times 10^{-4}$, which should be fairly reliable, is
obtained by comparing the dominant diagram for $B_s\to D_s^- K^+$, shown in
Fig.~2, to the very similar one for the decay $B_d\to D^-\pi^+$, whose
branching ratio is known. The branching ratio for the decay $B_s\to D_s^+
K^-$ (again of a pure $B_s$ neglecting mixing) is estimated, both in
Ref.~\bsdkref\ and by the present authors, to be $\sim 1\times 10^{-4}$.
This estimate is obtained by comparing the dominant diagram for $B_s\to
D_s^+ K^-$, shown in Fig.~3, to the related but somewhat different diagram
for $B^-\to\Psi K^-$. Accordingly, it is not as reliable as the estimate
for $B_s\to D_s^- K^+$.

\begin{figure}
\vspace{50mm}
\caption{The dominant diagram for $B_s\to D_s^- K^+$.}
\end{figure}

\begin{figure}
\vspace{50mm}
\caption{The dominant diagram for $B_s\to D_s^+ K^-$.}
\end{figure}

In $B_s(t)\to D_s^+ K^-$, the interfering decay amplitudes are, of course,
$A(B_s\to D_s^+ K^-)$ and $A(\bbar_s\to D_s^+ K^-)$. The amplitude
$A(B_s\to D_s^+ K^-)$, being dominated by the diagram of Fig.~3, has a CKM
phase which is $arg(V_{ub}^* V_{cs})\simeq \gamma$. This amplitude receives
no other tree-level contributions except from a $W$-exchange diagram with
the same CKM phase. Penguin diagrams cannot contribute at all. The
amplitude $A(\bbar_s\to D_s^+ K^-)$ is dominated by the diagram which is
the CP-mirror-image of that in Fig.~2. Thus, it has a CKM phase which is
$arg(V_{cb}V_{us}^*)\simeq 0$. It receives no other tree-level
contributions except from a $W$-exchange diagram with the same CKM phase,
and no penguin contributions. Thus, from Eq.~(\ref{CKMphase}), the CKM
phase $\varphi_{s,D_s^+ K^-}$ probed by the rates (\ref{nonCPrates}) for
the decays $B_s(t),\bbar_s(t)\to D_s^\pm K^\mp$ is $\gamma$. Moreover, from
our branching ratio estimates, the magnitudes $M_{s,D_s^+ K^-} = \vert
A(B_s\to D_s^+ K^-)\vert$ and ${\overline M}_{s,D_s^+ K^-} = \vert
A(\bbar_s\to D_s^+ K^-)\vert~ (=\vert A(B_s\to D_s^- K^+)\vert)$ of the two
interfering decay amplitudes in $B_s(t)\to D_s^+ K^-$ are in the ratio
$(1\times 10^{-4} / 2 \times 10^{-4})^{1/2} \simeq 0.7$. Thus, the desire
that these magnitudes be comparable is very nicely satisfied.

\noindent
$\bullet$ $B_s(t),\bbar_s(t)\to D^0\phi,\dbar\phi$:

Once again, we have a final state, $f=D^0 \phi$, which is distinct from its
CP conjugate, $\fbar = \dbar \phi$, and we use the expressions
(\ref{nonCPrates}) to analyze the four time-dependent decays $B_s(t)\to
D^0\phi$, $B_s(t)\to \dbar\phi$, $\bbar_s(t)\to D^0\phi$, and
$\bbar_s(t)\to \dbar\phi$. Tagging of the parent $B$ is
essential.

In $B_s(t)\to D^0 \phi$, the interfering decay processes are $B_s\to
D^0\phi$, which is dominated by the tree diagram in Fig.~4, and $\bbar_s\to
D^0\phi$, which is dominated by the tree diagram in Fig.~5. Penguin
diagrams cannot contribute. Thus, the CKM phase $\varphi_{s, D^0\phi}$
probed by $B_s(t)\to D^0\phi$ and the related decays is $\gamma$.

\begin{figure}
\vspace{50mm}
\caption{The dominant diagram for $B_s\to D^0\phi$.}
\end{figure}

\begin{figure}
\vspace{50mm}
\caption{The dominant diagram for $\bbar_s\to D^0\phi$.}
\end{figure}

The branching ratio estimate quoted in Table 1, $2\times 10^{-5}$, is for
$B_s\to\dbar\phi$, or for its CP-mirror-image $\bbar_s\to
D^0\phi$, and is obtained by comparing this process to $B_d\to\Psi K^{*0}$.
The diagram which dominates $B_s\to D^0\phi$ is almost identical to that
which dominates $\bbar_s\to D^0\phi$, apart from CKM factors, and we
estimate the branching ratio for $B_s\to D^0\phi$ to be $4\times 10^{-6}$.
The magnitudes $M_{s,D^0\phi} = \vert A(B_s\to D^0\phi)\vert$ and
${\overline M}_{s,D^0\phi} = \vert A(\bbar_s\to D^0 \phi)\vert$ of the two
interfering decay amplitudes in $B_s(t)\to D^0\phi$ then have the ratio
$(4\times 10^{-6} / 2 \times 10^{-5})^{1/2} \simeq 0.4$, which is $O(1)$,
as desired.$^\bsdphiref$

\noindent
$\bullet$ $B_s(t),\bbar_s(t)\to \rho^0 \ks$:

This mode, oft-proposed as a probe of $\gamma$, has the advantage of
yielding a CP eigenstate, so that the analysis is simplified. However, the
estimated branching ratio, obtained by comparing $B_s\to \rho^0\ks$ to
$B_d\to\Psi\ks$, is very small.

The decay amplitudes $A(B_s\to\rho^0\ks)$ and $A(\bbar_s\to\rho^0\ks)$ that
interfere in $B_s(t)\to \rho^0\ks$ are dominated, respectively, by the tree
diagram in Fig.~6 and by its CP-mirror-image. Thus, from
Eq.~(\ref{CKMphase}), the CKM phase $\varphi_{s,\rho^0\ks}$ probed by
$B_s(t),\bbar_s(t)\to\rho^0\ks$ via Eqs.~(\ref{CPrates}) is $2\gamma$.
Furthermore, as in all decays to a CP eigenstate, if one diagram dominates
$A(B_q\to f)$ and its CP-mirror-image dominates $A(\bbar_q\to f)$, these
two interfering decay amplitudes are of identical size. However, unlike the
other modes in Table 1, $B_s(t),\bbar_s(t)\to\rho^0\ks$ does involve
penguin contributions. Possibly, these are significant, and some of them
have CKM phases other than $\gamma$. Thus, in addition to having a small
branching ratio, $B_s(t),\bbar_s(t)\to\rho^0\ks$ may not be a clean probe
of $\gamma$.

\begin{figure}
\vspace{50mm}
\caption{The dominant diagram for $B_s\to \rho^0\ks$.}
\end{figure}

\ttwohead{2.3}{Non-$B_s$ Decay Modes That Can Probe Gamma}

While most decays of charged $B$ mesons cannot yield clean CKM phase
information, the decays $B^\pm\to DK^\pm$ are an exception, and they probe
$\gamma$.$^\GroWyler$ The technique for using these decays, explained in
Ref.~\GroWyler, requires one to measure the branching ratios for $B^\pm \to
D^0 K^\pm$, $B^\pm\to\dbar K^\pm$, and $B^\pm\to \dcp K^\pm$,
where $\dcp$ is a neutral $D$ that decays to a CP eigenstate such as
$K^+K^-$ or $\pi^+\pi^-$. As in all charged $B$ decays, there is, of
course, no need to tag, and no non-exponential time dependence.

The decay $B^+\to D^0 K^+$ is dominated by the diagram in Fig.~7, while
$B^+\to \dbar K^+$ is dominated by the diagram in Fig.~8. Since
$\dcp$ is a coherent superposition of $D^0$ and $\dbar$, in
$B^+\to \dcp K^+$ the diagrams of Figs.~7 and 8 interfere. Now, the CKM
phase of the $B^+\to D^0 K^+$ diagram, Fig.~7, is $arg(V_{ub}^*
V_{cs})\simeq \gamma$. That of the $B^+\to \dbar K^+$ diagram,
Fig.~8, is $arg(V_{cb}^* V_{us})\simeq 0$. Hence, the interference between
these diagrams probes $\gamma$. There are no penguin contributions, so this
probe is quite clean.

\begin{figure}
\vspace{50mm}
\caption{The dominant diagram for $B^+ \to D^0 K^+$.}
\end{figure}

\begin{figure}
\vspace{50mm}
\caption{The dominant diagram for $B^+ \to \dbar K^+$.}
\end{figure}

By comparing the diagram for $B^+ \to \dbar K^+$ to that for $B^+\to
\dbar\pi^+$, we readily estimate that $BR(B^+ \to \dbar K^+) \simeq 2
\times 10^{-4}$. This is a promising value. However, by comparing the
diagram for $B^+\to D^0 K^+$ to that for $B^+\to\Psi K^+$, we estimate that
$BR(B^+\to D^0 K^+) \sim 2 \times 10^{-6}$. In addition, by comparing
$B^+\to D^0 K^+$ to $B_d\to\dbar\pi^0$, for which there is an interesting
upper limit,$^\Besson$ we estimate that $BR(B^+\to D^0 K^+) \lsim
6\times 10^{-6}$. Thus, if these estimates prove to be right, this
branching ratio may be hard to measure. So may the branching ratios for
$B^\pm\to\dcp K^\pm$, since study of these processes requires that the
neutral $D$ decay to a CP eigenstate, a requirement which costs a factor
of $\sim 10^{-2}$ in overall branching ratio. The initial $B^\pm$ decay
will be dominated by the diagram of Fig.~8 or its CP-mirror-image, and so
will have a branching ratio of $\simeq 2\times 10^{-4}$. Thus, the overall
branching ratio will be $\sim 2\times 10^{-6}$.

As in all studies of CP violation in $B$ decay, one would like the two
diagrams which interfere in $B^+\to\dcp K^+$ to be of comparable magnitude.
{}From our branching ratio estimates, their magnitudes will be in the ratio
$\sim (2\times 10^{-6} / 2 \times 10^{-4})^{1/2} \simeq 1/10$. While not as
close to unity as one might wish, this ratio is perhaps close enough to
yield measurable interference effects.

A variant of the $B^\pm\to D K^\pm$ approach utilizing the self-tagging
$B_d$ decays $B_d(\bbar_d) \to D K^{*0}(D{\overline K}^{*0})$ has been
proposed as an alternate way to probe $\gamma$.$^\dunietz$ By comparing to
$B^+\to D^0 K^+$, we estimate that $BR(B_d\to D^0 K^{*0}) \sim 2\times
10^{-6}$, and by comparing to $B_d\to D^0 K^{*0}$, that $BR(B_d\to \dbar
K^{*0}) \sim 2\times 10^{-5}$.

\onehead{3.}{WHAT ANGLES DO THE ``GAMMA'' MODES}
\vskip-9truemm
\onehead{}{ACTUALLY MEASURE?}

We have identified a number of $B$ decay modes which, {\it within the
Wolfenstein approximation to the CKM matrix}, probe the angle $\gamma$. In
each of these modes, the decay amplitude consists of two interfering terms,
as illustrated in Eq.~(\ref{Binterference}), and each of these terms is
dominated by a single Feynman diagram. {\it In the Wolfenstein
approximation}, the CKM phases of these dominating diagrams are such that
the interfering terms in the decay amplitude have relative CKM phase
$\gamma$, or $2\gamma$. In this approximation, the statement that our
``$\gamma$ modes'' probe $\gamma$ entails only the error, which we have
argued is in most cases small or absent, corresponding to the neglect of
the non-dominating diagrams. However, suppose that one does not make the
Wolfenstein approximation. The CKM phases of Feynman diagrams are then
altered. Neglecting the non-dominating diagrams, do the ``$\gamma$ modes''
still probe $\gamma$? If not, what phase angle does each of them actually
probe? How big an error do we make if we identify this angle as being
approximately $\gamma$?

To explore these questions, we note that a very useful framework for
dealing with phases in the CKM matrix is provided by the ``unitarity
triangles.'' One of these triangles is shown in Fig.~1. When the CKM matrix
is treated exactly, rather than in the Wolfenstein approximation, one has,
not just this one triangle, but six triangles. These triangles correspond
to the unitarity constraint that any pair of columns, or any pair of rows,
of the CKM matrix be orthogonal. That is, they correspond to the
orthogonality requirements
\begin{eqnarray*}
& ds & ~~~~~~~ V_{ud}V_{us}^* + V_{cd}V_{cs}^* + V_{td}V_{ts}^* = 0 \\
& & ~~~~~~~~~~~ \lambda ~~~~~~~~~ \lambda ~~~~~~~~~~ \lambda^5
\end{eqnarray*}
\vskip-8truemm
\begin{eqnarray*}
& sb & ~~~~~~~ V_{us}V_{ub}^* + V_{cs}V_{cb}^* + V_{ts}V_{tb}^* = 0 \\
& & ~~~~~~~~~~~ \lambda^4 ~~~~~~~~ \lambda^2 ~~~~~~~~~ \lambda^2
\end{eqnarray*}
\vskip-8truemm
\begin{eqnarray*}
& db & ~~~~~~~ V_{ud}V_{ub}^* + V_{cd}V_{cb}^* + V_{td}V_{tb}^* = 0 \\
& & ~~~~~~~~~~~ \lambda^3 ~~~~~~~~ \lambda^3 ~~~~~~~~~ \lambda^3
\end{eqnarray*}
\vskip-8truemm
\begin{eqnarray}
\label{triconditions}
& uc & ~~~~~~~ V_{ud}V_{cd}^* + V_{us}V_{cs}^* + V_{ub}V_{cb}^* = 0 \\
& & ~~~~~~~~~~~ \lambda ~~~~~~~~~ \lambda ~~~~~~~~~~ \lambda^5 \nonumber
\end{eqnarray}
\vskip-8truemm
\begin{eqnarray*}
& ct & ~~~~~~~ V_{cd}V_{td}^* + V_{cs}V_{ts}^* + V_{cb}V_{tb}^* = 0 \\
& & ~~~~~~~~~~~ \lambda^4 ~~~~~~~~ \lambda^2 ~~~~~~~~~ \lambda^2
\end{eqnarray*}
\vskip-8truemm
\begin{eqnarray*}
& ut & ~~~~~~~ V_{ud}V_{td}^* + V_{us}V_{ts}^* + V_{ub}V_{tb}^* = 0 \\
& & ~~~~~~~~~~~ \lambda^3 ~~~~~~~~ \lambda^3 ~~~~~~~~~ \lambda^3
\end{eqnarray*}
To the left of each of these equations, we have indicated the pair of
columns, or pair of rows, whose orthogonality it expresses. Also, under
each term in the equations, we have indicated the rough magnitude of the
term as a power of the Wolfenstein parameter $\lambda=0.22$.

The unitarity triangles, depicted somewhat schematically in Fig.~9, are
simply pictures in the complex plane of the conditions
(\ref{triconditions}). Apart from signs and extra $\pi$'s, the angles in
any triangle are just the relative phases of the various terms in the
corresponding condition. Let us refer to a specific triangle by stating the
columns (rows) whose orthogonality it expresses, and a specific leg in this
triangle by stating which up-type (down-type) quark it involves.  Denoting
up-type quarks by Greek letters, and down-type ones by Latin letters, let
\beq
\omega_{\alpha\beta}^{ij}\equiv arg\left(V_{\alpha i}V^*_{\alpha j} /
V_{\beta i}V^*_{\beta j} \right)
\eeq
be the relative phase of the $\alpha$ and $\beta$ legs in the $ij$
triangle. Since
\beq
arg\left( V_{\alpha i} V^*_{\alpha j} / V_{\beta i} V^*_{\beta j} \right) =
arg\left( V_{\alpha i} V^*_{\beta i} / V_{\alpha j} V^*_{\beta j} \right),
\label{trianglerel}
\eeq
$\omega_{\alpha\beta}^{ij}$ is also the relative phase of the $i$ and $j$
legs in the $\alpha\beta$ triangle. That is, each angle in a triangle
expressing orthogonality of rows is also an angle in one expressing
orthogonality of columns. Hence, for our discussion of CKM phases, we can
forget about the row triangles. From the first two of
Eqs.~(\ref{triconditions}) (c.f.\ also Fig.~9), we see that
\begin{eqnarray}
\omega_{uc}^{ds} & \le & O(\lambda^4), \nonumber\\
\omega_{ct}^{sb} & \le & O(\lambda^2).
\label{smallangles}
\end{eqnarray}
That is, one of the angles in the $sb$ triangle is small ($\lsim 0.05$),
and one in the $ds$ triangle is extremely small ($\lsim 0.003$). (There is
no reason to suppose that the remaining angles in these triangles are
small.)

\begin{figure}
\vspace{150mm}
\caption{The unitarity triangles. To the left of each triangle is indicated
the pair of columns, or of rows, whose orthogonality this closed triangle
expresses.}
\end{figure}

Now, when the CKM matrix is treated exactly, what CKM phases do the decay
modes we have discussed in Sec.~2 actually probe? Any neutral $B$ mode
probes the phase given by Eq.~(\ref{CKMphase}). Applied to any $B_s(t)$
decay, this equation involves the mixing phase
$arg(B_s\to{\overline{B_s}})$, which from Eq.~(\ref{omegaq}) is
$arg(V_{ts}V_{tb}^*/V_{ts}^*V_{tb})$. Moreover, the $V_{ts}/V_{ts}^*$ in
this expression cannot be cancelled by the decay amplitudes $A(B_s\to f)$
and $A({\overline{B_s}}\to f)$ so long as these amplitudes are dominated by
tree diagrams, which can never involve a $t$ quark. However, from Fig.~1,
apart from a $\pi$,
\beq
\gamma \equiv \omega_{uc}^{db} = arg(V_{ud} V_{ub}^*/V_{cd} V_{cb}^*).
\eeq
Since the CKM elements in this expression do not include $V_{ts}$, it is
clear that the phase probed by a $B_s(t)$ decay cannot be $\gamma$.

For the decay $B_s(t)\to D_s^+ K^-$, $A(B_s\to f)$ is dominated by the
diagram in Fig.~3, proportional to $V_{cs} V_{ub}^*$. Similarly,
$A({\overline{B_s}}\to f)$ is dominated by the CP-mirror-image of the
diagram in Fig.~2, and so is proportional to $V_{cb} V_{us}^*$. Thus, from
Eqs.~(\ref{CKMphase}) and (\ref{omegaq}), the phase probed by $B_s(t)\to
D_s^+ K^-$ is
\begin{eqnarray}
\varphi_{s,D_s^+ K^-} & = & arg \left[ { V_{cs} V_{ub}^* \over
\displaystyle {\strut {V_{ts} V_{tb}^* \over V_{ts}^* V_{tb} } } \,
V_{cb} V_{us}^* } \right] \nonumber\\
& = & arg \left[ { V_{ud} V_{ub}^* \over V_{cd} V_{cb}^* }
\left( { V_{cs} V_{cb}^* \over V_{ts} V_{tb}^* } \right)^2
{ V_{us} V_{ud}^* \over V_{cs} V_{cd}^* } \right] \nonumber\\
& = & \gamma + 2 \, \omega^{sb}_{ct} - \omega_{uc}^{ds}~.
\end{eqnarray}
We see that this phase is $\gamma$ plus angles in the $sb$ and $ds$
triangles. From Eq.~(\ref{smallangles}), we note that the particular $sb$
and $ds$ angles involved are the small ones, so that $\varphi_{s,D_s^+
K^-}$ is $\gamma$ plus a $\le O(\lambda^2)$ correction.

In the same way, we can find the CKM phases probed by the other $B_s(t)$
decay modes listed in Table 1. For the decays
\begin{eqnarray}
\label{seqBDK}
B^\pm & \to & D + K^\pm~, \\
& & ~~~~~~f_\cp \nonumber
\end{eqnarray}
where $f_\cp$ is the CP eigenstate (e.g.\ $K^+K^-$) into which the
neutral $D$ decays, we must find the relative CKM phase of the two
interfering terms in the decay amplitude
\begin{eqnarray}
A\left(\matrix{B^+\to D + K^+ \cr ~~~~~~~~~~~~~f_\cp \cr} \right)
& = & A(B^+\to D^0 K^+) A(D^0\to f_\cp) \nonumber\\
& & \qquad\qquad\qquad
+ A(B^+\to {\overline{D^0}} K^+) A({\overline{D^0}}\to f_\cp).
\label{BDKamplitude}
\end{eqnarray}
If $D^0$-${\overline{D^0}}$ mixing is slow compared to the $D^0$ decay
rate, then, as suggested by Eq.~(\ref{BDKamplitude}), the $D$-system phases
which influence the decay sequence (\ref{seqBDK}) are the $D$ {\it decay}
phases, not the $D^0$-${\overline{D^0}}$ mixing phase. But then the phase
probed by the sequence depends on $f_\cp$. For $f_\cp=K^+ K^-$, we find
from the diagrams of Figs.~7 and 8, and the tree diagrams for $D^0\to
K^+K^-$ and ${\overline{D^0}}\to K^+K^-$, that the relative CKM phase of
the two terms in Eq.~(\ref{BDKamplitude}) is
\begin{eqnarray}
arg\left[ V_{us} V_{ub}^* / V_{cs} V_{cb}^* \right] & = & \omega_{uc}^{sb}
\nonumber\\
& = & arg \left[ { V_{ud} V_{ub}^* \over V_{cd} V_{cb}^* }
{ V_{us} V_{ud}^* \over V_{cs} V_{cd}^* } \right] =
\gamma - \omega_{uc}^{ds}~.
\end{eqnarray}
Thus, the decay chain (\ref{seqBDK}) with $f_\cp=K^+ K^-$ probes a CKM phase
which is one of the ``large'' angles in the $sb$ triangle, and this angle
is in turn $\gamma$ plus a $\le O(\lambda^4)$ correction.

\begin{table}
\hfil
\vbox{\offinterlineskip
\halign{&\vrule#&
   \strut\null~\hfil#\hfil\null~\cr
\noalign{\hrule}
height2pt&\omit&&\omit&\cr
& Decay Mode && CKM Phase Probed & \cr
height2pt&\omit&&\omit&\cr
\noalign{\hrule}
height5pt&\omit&&\omit&\cr
& $B_s,\bbar_s \to D_s^\pm K^\mp$ && $\gamma + 2\omega_{ct}^{sb} -
\omega_{uc}^{ds}$ & \cr
height5pt&\omit&&\omit&\cr
& $B_s,\bbar_s \to D^0\phi,\dbar\phi$ && $\gamma + 2\omega_{ct}^{sb} -
\omega_{uc}^{ds}$ & \cr
height5pt&\omit&&\omit&\cr
& $B_s,\bbar_s \to \rho^0 \ks$ && $2(\gamma+\omega_{ct}^{sb})$ & \cr
height5pt&\omit&&\omit&\cr
& $B^\pm \to D K^\pm$ [with $D\to K^+K^-$] && $\gamma-\omega_{uc}^{ds}$ &
\cr
height5pt&\omit&&\omit&\cr
& $B_d (\bbar_d) \to D K^{*0} (D{\overline K}^{*0})$ [with $D\to K^+K^-$]
&& $\gamma-\omega_{uc}^{ds}$ & \cr
height5pt&\omit&&\omit&\cr
\noalign{\hrule}}}
\caption{CKM phases probed by the ``$\gamma$ modes'' when the CKM matrix is
treated exactly.}
\end{table}

In Table 2 we show what CKM phases are actually probed by the various
``$\gamma$ modes'' we have considered when the CKM matrix is treated
exactly. These phases are expressed in terms of $\gamma$ and angles in the
$sb$ and $ds$ triangles. We see from Table 2 that {\it none} of the modes
we have discussed actually measures $\gamma$. Every one of them yields
$\gamma$, or $2\gamma$, plus nonzero corrections. On the other hand, in
every case the corrections involve only the $\le O(\lambda^2)$ angle in the
$sb$ triangle and/or the $\le O(\lambda^4)$ angle in the $ds$ triangle.
Thus, the corrections are always less than $0.1$ radians. One might wonder
whether the correction angles $\omega_{ct}^{sb}$ and $\omega_{uc}^{ds}$
can represent a {\it fractionally} large correction in the event that
$\gamma$ itself is small. It can be shown that they cannot. When $\gamma$
goes to zero, $\omega_{ct}^{sb}$ and $\omega_{uc}^{ds}$ also go to zero, at
the same rate as $\gamma$. Furthermore, given what is already known about
the CKM matrix, the proportionality constant relating $\omega_{ct}^{sb}$ to
$\gamma$ for small $\gamma$ is $\sim 0.015$, and that relating
$\omega_{uc}^{ds}$ to $\gamma$ is still smaller. Thus, the corrections to
$\gamma$ are always fractionally small.

As the examples in Table 2 suggest, {\it any} $B$ decay mode which probes
$\gamma$ in the Wolfenstein approximation probes $\gamma$ plus, at most,
corrections involving only the small angles $\omega_{ct}^{sb}$ and
$\omega_{uc}^{ds}$ when the CKM matrix is treated exactly. For, as
discussed in Sec.~4, the exact CKM phase probed by an arbitrary $B$ decay
mode can be expressed as a linear combination of $\gamma$, $\beta$,
$\omega_{ct}^{sb}$ and $\omega_{uc}^{ds}$, with integer coefficients. Now,
for a $B$ decay which yields $\gamma$ in the Wolfenstein approximation,
this linear combination obviously does not involve $\beta$. Thus, apart
from $\gamma$, it can involve only $\omega_{ct}^{sb}$ and
$\omega_{uc}^{ds}$.

While the angles measured by the various ``$\gamma$ modes'' differ only
slightly from $\gamma$, they do differ, and in some modes they may differ
by as much as 0.05 to 0.1. In contrast, neglecting penguin contributions,
the ``$\alpha$ modes'' $B_d,\bbar_d\to\pi^+ \pi^-$ and $B_d,\bbar_d \to
\rho^\pm \pi^\mp$ yield precisely $\alpha$, even when the CKM matrix is
treated exactly. Similarly, assuming as usual that $K^0$-${\overline{K^0}}$
mixing is dominated by the $d{\bar s}\to c{\bar c} \to s{\bar d}$ box
diagram, the ``$\beta$ mode'' $B_d,\bbar_d\to\Psi\ks$ yields precisely
$\beta$. Thus, in the second generation experiments on CP violation in $B$
decays, it would be interesting to test the Standard Model by showing that
the angles extracted from, say, the modes $B_d,\bbar_d\to\pi^+\pi^-$,
$B_d,\bbar_d\to\Psi\ks$ and $B_s,\bbar_s\to D_s^\pm K^\mp$ {\it fail} to
add up to $\pi$ by an amount of order 0.05 to 0.1. To carry out this
precision test of the angles probed by the leading diagrams in various
modes, one would have to eliminate from $B_d,\bbar_d\to\pi^+\pi^-$ the
possible penguin contributions, using an isospin analysis.$^\isospin$

\begin{figure}
\vspace{100mm}
\caption{Diagrams for $B_c^+\to D^0\pi^+$.}
\end{figure}

It is tempting to ask whether there is {\it any} $B$ decay mode which,
unlike all the modes we have discussed, actually measures precisely
$\gamma$ when the CKM matrix is not approximated. In principle, the mode
$B_c^+\to D^0\pi^+$ does this via interference between the diagrams shown
in Fig.~10. The relative CKM phase of these diagrams is just
$arg(V_{ud}V_{ub}^*/V_{cd}V_{cb}^*)=\gamma$. However, it is not clear that
the penguin diagrams for this decay are small relative to the annihilation
diagram in Fig.~10,$^\isicomment$ and, in any case, this mode, like most
charged $B$ decays, cannot yield clean CKM phase information.

It would, of course, be very interesting to probe directly angles in the
$sb$ and $ds$ triangles. One decay mode that would do this is
$B_s(t),\bbar_s(t) \to \Psi\phi$.$^\Psiphi$ In this mode, the interfering
decay amplitudes are $A(B_s\to\Psi\phi)$,which is dominated by the diagram
in Fig.~11, and $A(\bbar_s\to\Psi\phi)$, which is dominated by its CP
conjugate. From Eqs.~(\ref{CKMphase}) and (\ref{omegaq}), the CKM phase
probed by $B_s(t)\to\Psi\phi$ is then$^\Gronauhelp$
\begin{eqnarray}
\varphi_{s,\Psi\phi} & = & arg \left[ { V_{cs} V_{cb}^* \over
\displaystyle {\strut {V_{ts} V_{tb}^* \over V_{ts}^* V_{tb} } } \,
V_{cb} V_{cs}^* } \right] \nonumber\\
& = & 2 \, arg \left[ { V_{cs} V_{cb}^* \over V_{ts} V_{tb}^* } \right]
= 2 \, \omega_{ct}^{sb}~,
\label{bigsmallangle}
\end{eqnarray}
just twice the small angle in the $sb$ triangle.

\begin{figure}
\vspace{50mm}
\caption{The dominant diagram for $B_s\to \Psi\phi$.}
\end{figure}

To use this mode, tagging and measurement of the time dependence will, of
course, be necessary. The final state $\Psi\phi$ is technically not a CP
eigenstate, because it may be a mixture of helicity configurations.
However, it appears that the outgoing particles in the decay $B_d\to\Psi
K^*$ have zero helicity much of the time.$^\PsiKstar$ We then expect the
same to be true of the outgoing particles in
$B_s(t),\bbar_s(t)\to\Psi\phi$. The final state then {\it is} largely a CP
eigenstate, and the decay rates are described by the simple equations
(\ref{CPrates}).

By comparing to the very similar decay $B_d\to\Psi K^*$, one readily
estimates that $BR(B_s\to\Psi\phi)\simeq 10^{-3}$, a very promising value.

It has been pointed out$^\Psiphi$ that $\sin\varphi_{s,\Psi\phi}$, the
quantity probed by $B_s(t),\bbar_s(t)\to\Psi\phi$ via Eqs.~(\ref{CPrates}),
can be rewritten as
\beq
\vert \sin\varphi_{s,\Psi\phi} \vert = 2 \, \left\vert V_{cd}
{\displaystyle V_{ub}\over V_{cb} } \sin\gamma \right\vert
(1+O(\lambda^2)).
\label{Psiphiasym}
\eeq
Thus, if $\vert V_{ub}/V_{cb} \vert$ is known, this decay mode becomes
another way to determine $\gamma$. Of course, the mode does not ``probe
$\gamma$'' in the sense of involving two interfering amplitudes whose
relative CKM phase is $\gamma$ or $2\gamma$. Rather, the relative phase is,
as we saw in Eq.~(\ref{bigsmallangle}), the small phase
$2 \, \omega_{ct}^{sb}$. The CP-violating asymmetry in the mode will be
correspondingly small, rather than being of order $\sin\gamma$ or $\sin
2\gamma$. Indeed, if we assume that $\vert V_{ub}/V_{cb} \vert \sim
0.07$,$^\Besson$ then Eqs.~(\ref{Psiphiasym}) and (\ref{CPrates}) indicate
that the asymmetry will be $\sim 0.03\sin\gamma$, which is necessarily
quite small. Nevertheless, perhaps the large branching ratio for the mode
will make this small asymmetry observable. Needless to add, if
$B_s(t),\bbar_s(t)\to\Psi\phi$ should be found to have an asymmetry much
larger than, say, 0.06, then we would have evidence for a CP-violating
mechanism beyond that in the Standard Model.

\onehead{4.}{THE UNITARITY TRIANGLES AND THE CKM MATRIX}

The discussion of the previous Section calls attention to the unitarity
triangles beyond the $db$ triangle, and to the angles in those other
triangles. We would now like to briefly report the results of a general
analysis of how the angles in the full set of six unitarity triangles are
related to CP-violating asymmetries in $B$ decay, and of how they are
related to the CKM matrix. A more complete discussion, including the proofs
of the results, will be presented elsewhere.$^\AKL$

There are three main results, which we shall describe in turn.

\begin{itemize}

\item As we have already noted (see Eq.~\ref{trianglerel}), each angle in a
``row'' triangle is also an angle in a ``column'' triangle. Thus, there are
at most nine, not eighteen, distinct angles in the six unitarity triangles.
We find that exactly four of these nine angles are independent. The
remaining five angles are very simple linear combinations of the
independent four. The independent angles may be chosen, for example, as two
of the angles $\alpha$, $\beta$ and $\gamma$ in the $db$ triangle, plus two
of the angles in one of the other column triangles. They may also be chosen
as two of the angles $\alpha$, $\beta$ and $\gamma$, plus the two small
angles $\omega_{ct}^{sb}$ and $\omega_{uc}^{ds}$.

\item As we have seen, any CP-violating asymmetry in $B$ decay probes the
relative CKM phase of two interfering amplitudes. Thus, the asymmetry
probes the phase of some product and quotient, or, equivalently, of some
product, of CKM elements. Now, not every conceivable product of CKM
elements has a phase which is invariant under phase redefinitions of the
quark fields. However, if the phase of some product of CKM elements is
determined by an experiment, then, obviously, this phase must be invariant
under quark-field rephasing.

We find that if the phase $\varphi$ of some product of CKM elements
is rephasing-invariant, then
\beq
\varphi = \sum_{i=1}^4 n_i\varphi_i + k\pi~,
\label{anglebasis}
\eeq
where the $\varphi_i$ are the four independent angles in the unitarity
triangles, the $n_i$ are integers, and $k$ is zero or one. For any specific
$\varphi$ and choice of the independent angles $\varphi_i$, the $n_i$ and
$k$ will, of course, be known quantities.

The relation (\ref{anglebasis}) states that the CKM phase probed by any
CP-violating asymmetry in a $B$ decay is a simple linear combination, with
integer coefficients, of the four independent angles in the unitarity
triangles. Thus, these angles form a complete set of variables for the
description of CP violation in $B$ decay. Moreover, these variables are
very closely and simply related to the quantities -- the phases $\varphi$
-- to be measured in $B$ decay experiments.

\item Suppose that the four independent angles $\varphi_i$ have been
determined by CP-violation experiments. What have we learned about the CKM
matrix? The answer is that once the $\varphi_i$ are known, the entire CKM
matrix follows from them! That is, in principle at least, we can determine
the whole CKM matrix, including the magnitudes of all its elements, and all
its physically-meaningful phases, through measurements of nothing but
CP-violating asymmetries in $B$ decays. Thus, albeit at varying levels of
sensitivity, these CP-violating asymmetries probe {\it everything} in the
CKM matrix. Consequently, they are potentially a very rich test of the
Standard Model explanation of CP violation.

\end{itemize}

\onehead{}{ACKNOWLEDGEMENTS}

It is a pleasure to thank A. Ali, I. Dunietz, M. Gronau, J. Rosner and P.
Sphicas for helpful conversations. We thank the organizers for their
thoughtful, supportive and gracious efforts, which made possible a very
fruitful Workshop. Two of us (BK and DL) are grateful for the hospitality
of Fermilab, where part of this work was done. One of us (BK) is also
grateful for the hospitality of DESY, where another part was done, and for
that of the Universit\'e de Montr\'eal, where the paper was completed. This
work was supported in part by the SSC Laboratory, the Natural Sciences and
Engineering Research Council of Canada, and by FCAR, Qu\'ebec.


\end{document}